\documentclass[prl,twocolumn,showpacs,preprintnumbers,amsmath,amssymb]{revtex4}
\usepackage{graphicx}% Include figure files
\usepackage{dcolumn}% Align table columns on decimal point
\usepackage{bm}% bold math
\usepackage{psfig}
\newcommand \be{\begin{eqnarray}}
\newcommand \ee{\end{eqnarray}}
\begin{document}
\title{
%Charge ratchets: 
Current without bias and diode effect in shuttling transport of nanoshafts} 
\author{K. Morawetz$^{1,2}$, S. Gemming$^{1}$, R. Luschtinetz$^{3}$, L. M. Eng$^{4}$,  G. Seifert$^{3}$, A. Kenfack$^{2}$
%,   P. Lipavsk\'y$^{5,6}$
}
\affiliation{$^1$Forschungszentrum Dresden-Rossendorf, PF 51 01 19, 
01314 Dresden, 
Germany}
\affiliation{$^2$Max-Planck-Institute for the Physics of Complex
Systems, N\"othnitzer Str. 38, 01187 Dresden, Germany}
\affiliation{$^3$Institute of Physical Chemistry and Electrochemistry, TU Dresden, 01062 Dresden, Germany}
\affiliation{$^4$Institute of Applied Photophysics, TU Dresden, 01062 Dresden, Germany}
%\affiliation{
%$^5$Faculty of Mathematics and Physics, Charles University, Ke Karlovu 3, 12116 Prague 2, Czech Republic}
%\affiliation{
%$^6$Institute of Physics, Academy of Sciences, Cukrovarnick\'a 10, 16253 Prague 6, Czech Republic
%}
\begin{abstract}
A row of parallely ordered and coupled molecular nanoshafts is shown to develop a shuttling transport of charges at finite temperature. The appearance of a current without applying an external bias voltage is reported as well as a natural diode effect allowing unidirectional charge transport along one field direction while blocking the opposite direction. The zero-bias voltage current appears above a threshold of initial thermal and/or dislocation energy. 
\end{abstract}
\maketitle

Organic field effect transistors (OFETs) based on different polymers \cite{YT05,KMHSTWF06,HJAKLEGHS08} attract an increased interest due to numerous potential applications as flexible and low cost storage and microelectronic devices.
% based on established OLED technologies.
For achieving excellent electric properties such as high charge carrier mobilities and low resistive losses, which are required for technologically attractive device applications, a high structural ordering of the semiconductor molecular material is necessary \cite{DM02,Ga99}. Oligothiophenes and their derivatives can be regarded as one of the most promising systems for building such self-organized structures across multiple length scales due to the variety of intra- and intermolecular interactions, which originates from the polarizability of the sulfur electrons and the aromatic $\pi$-electron system \cite{MGBCBMO03,MGTB00,BZBA93}. Just recently an OFET structure has been built from ultra-thin self-assembly films made up from 
%$\alpha$, $\omega$- and $\beta$, $\beta'$-substituted 
oligothiophenes, which are arranged in a high-order lamellar stacking perpendicular to the substrate surface \cite{HJAKLEGHS08}. 

In general, the charge transport is largest in the direction perpendicular to the plane of the thiophene rings \cite{F07,YDJIC02}. This finding addresses already the basic difference between electronic transport in organic conductors and in classical semiconductors. Though the band gap between the lowest unoccupied and the highest occupied level of about $3$ eV in quarterthiophene suggests an analogy to the conduction and valence band in semiconductors, there are crucial differences. While in conventional semiconductors the transport is due to delocalized states and limited by the scattering, the transport in molecules is due to localized states dominated by hopping \cite{H98}. 

In this paper we suggest a new mechanism of charge transport which is possible for flexible molecular tubes, shafts or any elastically deformable assembly which could provide an alternative explanation of transport properties in nanoshaft-based OFETs. The performance of OFETs based on oligothiophene \cite{HJAKLEGHS08} shows characteristic features, e.g. the current starts at a certain threshold of gate voltage and reaches a saturation value for certain drain voltages. 
We will show in the following that for a regular arrangement of long 
elastic molecules a shuttling transport can be established, which can 
model these features. In particular, we will describe a shuttling effect 
of nanoshafts leading to a diode behavior as well as to a directed current without external bias voltage. This resembles the ratchet effect described as Brownian motors \cite{HB96,JAP97,AH02} also realizable with oscillating laser fields \cite{LKMH04,KH07}. The thermal noise plays an important role for activating such motors to overcome a certain barrier \cite{HTB90}. In the effect presented here the activating threshold is given by an initial kinetic or potential energy necessary for the first nanoshaft to reach the contacts.

\begin{figure*}[t]
\centerline{\parbox[t]{16.3cm}{
\psfig{file=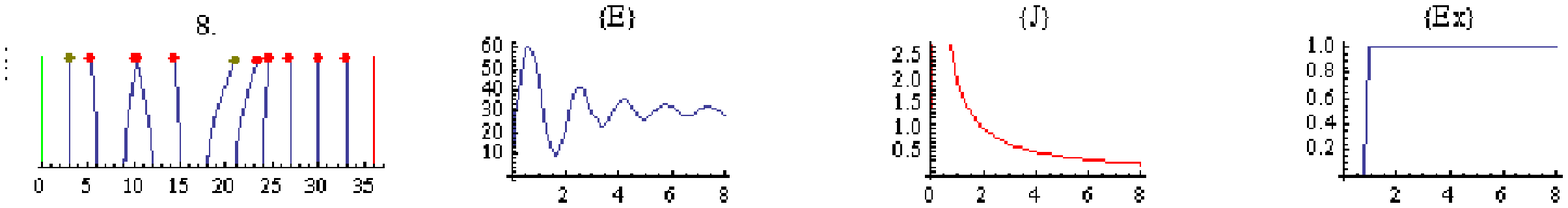,width=16cm}
\psfig{file=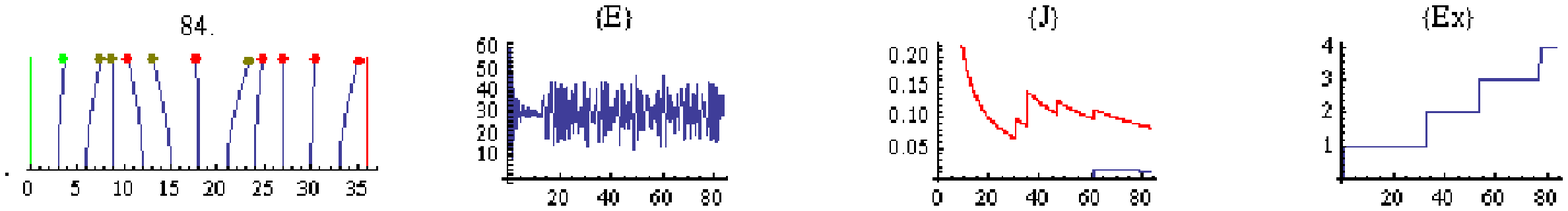,width=16cm}
\psfig{file=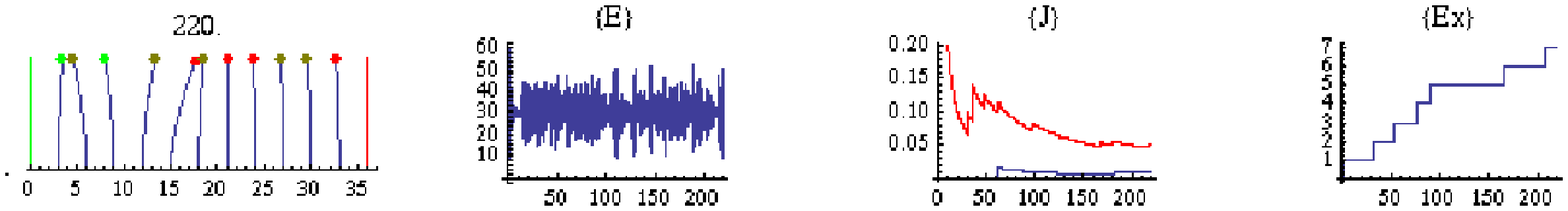,width=16cm}
}}
\vspace*{-3ex}
\caption{Snapshots of time evolution of the chain of shuttling nanoshafts (left)  with positive charge (green) and negative charges (red) or no charge (grey). The kinetic energy, currents on both sides and the recombination rate are plotted from left to right.\label{kette}}
\end{figure*}

Shuttling in single-dot devices had been already in the center of interest \cite{DNJ03,DNJ05}. Besides promising applications for nano-electromechanical devices the coupling of quantized transport with micromechanical cantilevers \cite{FNJ05} bridges the classical mechanical and the quantum physics in an exciting 
way. Our model, analytically solvable, demonstrates that a similar shuttling mode as known from single-dot transport by mechanical cantilevers is achievable by a row of elastically coupled molecular tubes. The model is based on the coupled spring chain. Despite their simplicity, such models surprise with the ability to describe unknown phenomena of classical quantization phenomena in velocities \cite{VMDST06}. The model of freely rotating linear chains has been investigated also in view of quantization \cite{E92} and has been applied to entanglement problems \cite{PKPL05}. We note that the shuttling transport described here should not be mixed up with the so called `chain shuttling' in polymerization reactions, where a growing polymer chain is transferred between different metal catalysts \cite{ACHKW06,ABBCS07}.  

The elasticity of polymers with respect to the bending and rigidity has been studied by molecular dynamic simulations \cite{BG05}. The model of hard-sphere chains is used to describe the stiffness and  diffusion in polymers \cite{BCRC04}. These elastic properties play an important role in the mesoscale modeling of carbon nanotubes \cite{B06}. Such carbon nanotubes on glassy carbon electrodes can enhance the current response by a factor of 1000 as reported in \cite{EBWB05}. Therefore it is of high actual interest to reinvestigate the transport in nanotubes. We will concentrate in this paper on charge transport perpendicular to the direction of nanoshafts.

Specifically, we consider $N$ nanoshafts of length $L$ parallely oriented in the $y$-direction. The shafts are standing perpendicular on the substrate, and are periodically arranged along the $x$-direction at a certain distance from each other (see Fig. \ref{kette}). We expect them to couple elastically by their top ends. Then the shafts bend in x direction when applying the tangential force $F$ with the displacement
\be
x(z)={F z^2 \over 2 {\cal E} I} \left (L-\frac z 3 \right )=\frac 3 2 x(L) \left ({z\over L}\right )^2 \left (1 - {z\over 3 L} \right )
\label{xe}
\ee 
where ${\cal E}$ is the Young modulus and $I=\int z^2 dA$ is the area moment of inertia being $I={\pi} r_0^4/4$ for circular cross sections of tubes with radius $r_0$ or $I= a b^3/12$ for nanoshafts with rectangular cross section of sides $a$ and $b$. The maximal displacement on the top ends $z=L$ is $x \equiv x(l)={F} L^3/3 {\cal E} I$ from which the spring constant $k=F/x=3 {\cal E} I/L^3$ is given provided we know the Young modulus. In other words we have a coupled linear chain of top ends obeying the differential equations ${d^2\over d t^2}x_i= \omega^2 (x_{i-1}-2 x_i+x_{i+1})$ with $\omega^2= k/m$ given by the mass of the nanoshafts. The nanoshafts are located in-between two solid contacts such that the coupled equations are bounded which can be formally expressed by $x_0=x_{N+1}=0$. Additionally each nanoshaft can carry a charge $q_i$ which can be exposed to a time-dependent external field $E(t)$. The total coupled equation system for the time-dependence of the nanoshaft top ends reads
\be
{d^2 x_i(t)\over d t^2}= \omega^2 [x_{i-1}(t)-2 x_i(t)+x_{i+1}(t)] +a_i(t)
\label{sys}
\ee
with $\omega^2=3 {\cal E} I/mL^3$
and $a_i(t)=q_i E(t)/m$. For simplicity we will use the time in units of $1/\omega$ further on. For quarterthiophene we have a typical value of $k=1$N/m and a mass of $m=5.49\times 10^{-22}$g which leads to the scale $\omega=1.35$/ps. The length will be in units of $l_0=0.1$nm as typical value. The field is given in units of $m \omega^2 l_0/q=6.2\times 10^8$V/m again given explicitly for quarterthiophene and the charge current in units of $q \omega=21.6$mA where we assume one single charge $q=e$. The typical energy units are $m \omega^2 l_0^2/2=31.2$meV$=362$K.

The analytical solution of (\ref{sys}) obtained by the standard eigenvalue method is given in terms of the 
normalized orthogonal system $\phi_{n\nu}=\sqrt{2/(N+1)} \sin{[n \nu \pi/(N+1)]}$,
\be
x_i(t)&=&\sum\limits_{n=1}^N \phi_{n i}\left. \biggl [ (c_n \cos{\omega_n t}+d_n \sin{\omega_n t})\right .
\nonumber\\&&\left .
+\int\limits_0^t dt' {\sin{\omega_n (t-t')}\over \omega_n} \sum\limits_{m=1}^N a_m(t') \phi_{nm} \right ]
\label{sol}
\ee
with the eigenfrequencies $\omega_n^2 =2(1-\cos{[n \pi/(N+1)]})$ and the initial condition determining $c_n=\sum_\nu \phi_{n\nu} x_\nu(0)$ and $d_n=\sum_\nu \phi_{n\nu} \dot x_\nu(0)/\omega_n$.

Now we proceed by considering the nanoshafts between two oppositely charged plates, $q_{\rm sides}$, modeling the contacts. Each nanoshaft can carry a negative charge $q_i<0$ describing the charge in the lowest unoccupied molecular orbital or a positive charge $q_i>0$ for the transport of holes in the highest occupied molecular orbital or none. Each time when two top ends of the nanoshafts touch each other the charge is moved if one of both tubes had no charge. In the case of opposite charges they annihilate and it is counted as recombination which gives rise to light emission. If one nanoshaft is touching the contact on one side it contributes to the current on this side as $q_{\rm sides}-q_i$. Each time such an event happens the time evolution according to the analytical solution (\ref{sol}) restarts with the new initial conditions and new charge distribution. In this way we use the speed of analytical solution together with the nonlinear process of recharging.

For exploratory reasons we restrict ourselves to 11 nanoshafts of length $20 l_0$ and have arranged them at a distance of $3 l_0$. These are typical parameters of quarterthiophene used in recent experiments \cite{HJAKLEGHS08}. First we charge all shafts equivalently and start the process of shuttling by bending the most left tube to the left contact, allowing to transfer the first charge to the left lead. 

\begin{figure}[h]
\psfig{file=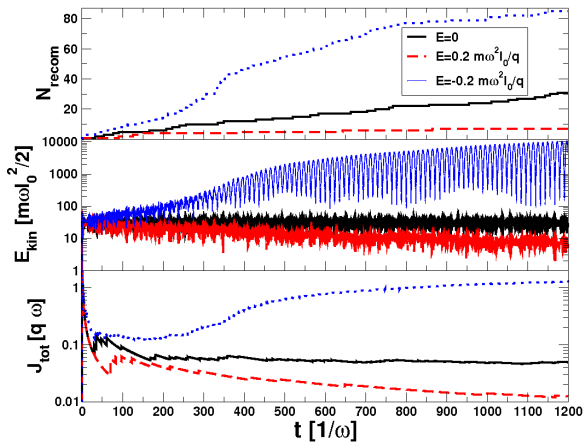,width=8cm}
\vspace*{-3ex}
\caption{The time evolution for the recombination rate (above), the total kinetic energy (middle) and the total current (below) for 3 different applied fields. The time unites are in $\omega^2=3 E I/mL^3$ and the length units are $l_0$. \label{shut_1}}
\end{figure}

In figure \ref{kette} we have plotted different snapshots of the time evolution in units of $1/\omega$. We see that the system starts to shuttle and to transport positive charges from the left to the right and negative charges from the right to the left side. This leads to steps in the current counted as charges delivered on the corresponding side per elapsed time. The recombination rate is increasing with time; complete movies can be found in \cite{mov}.

\begin{figure}[h]
\psfig{file=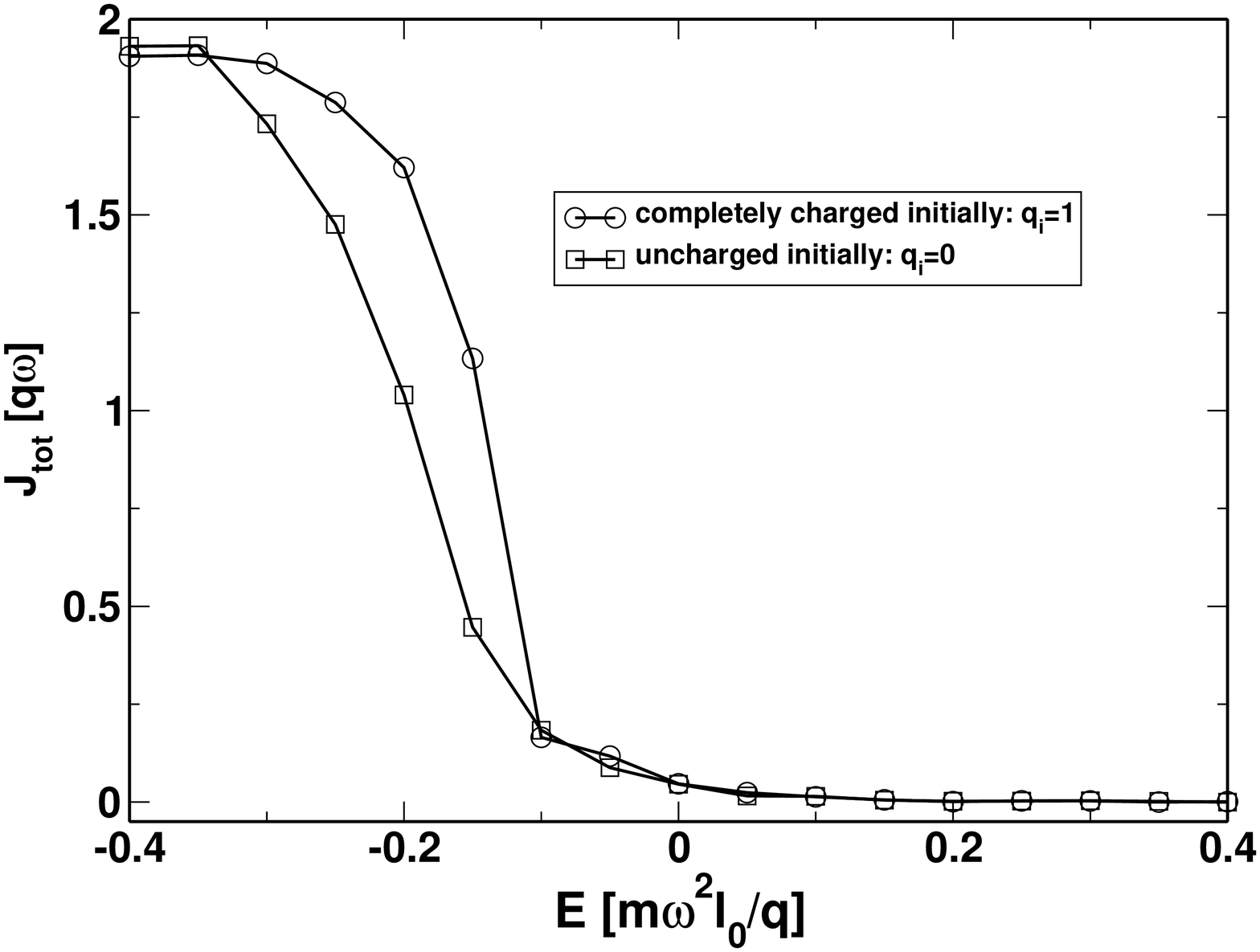,width=8cm}
\vspace*{-3ex}
\caption{The total current versus applied electric field for two cases, initially uncharged tubes and initially charged tubes. \label{feld}}
\end{figure}

The snapshots in figure \ref{kette} are actually taken from a run without applying any external field. The gates serves here merely as a reservoir of charges, not as capacitor plates. The astonishing observation is now that even for such an unbiased case a finite total current is developing.
This fact is demonstrated in figure \ref{shut_1} where the time evolution of the recombination rate, the kinetic energy, and the total current which is the sum of right and left currents, are plotted for different external fields. The total current and the recombination rate reach saturation with increasing time while the kinetic energy remains constant on average. If one applies an electric field of $E=0.2$ oppositely to the direction of zero bias voltage current we see that the kinetic energy is decreasing with time and the shuttling current as well as  the recombination rate are reduced. In fact, it reaches a state where the left and right half of the tubes shuttle symmetrically against each other such that no net current is delivered. If the field is applied in the opposite direction $E=-0.2$, the system is accelerated and delivering more and more charges, limited finally only by the length of the tubes. In other words we observe a pure diode effect just due to shuttling of coupled chains of nanoshafts. The total current reached after long times for different applied fields is depicted in figure \ref{feld} and illustrates this diode effect as a quite systematic one.

%\begin{figure}
%\psfig{file=shut_qd.eps,width=9cm}
%\caption{The time evolution of the total current for different applied fields. The time unites are in $\omega^2=3 E I/mL^3$, the field units are $m \omega^2/q$  and the length units are $l_0$. \label{shut_qd}}
%\end{figure}

One may argue that this diode effect as well as the observation of a current without external bias is due to the initial charging of the nanoshafts. In the figure \ref{feld} we have plotted the finally reached total current versus the applied field for two cases, the initially charged case considered so far, and the case where initially all shifts are uncharged. We see that identical final currents are obtained without external bias for both cases and the diode effect is present, as well.

\begin{figure}[h]
\psfig{file=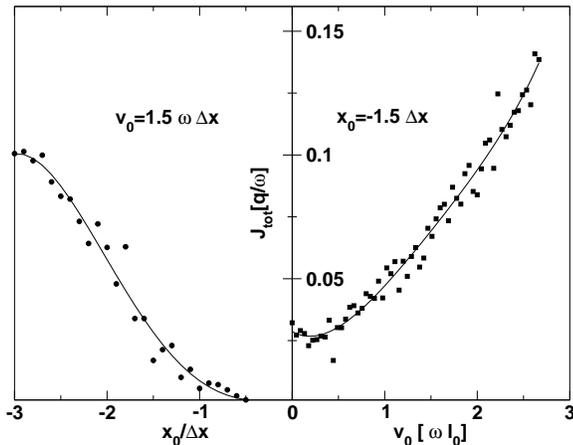,width=8cm}
\vspace*{-3ex}
\caption{The total current without voltage bias versus initial displacement (left) and versus initial velocity (right) of the most left nanoshaft.\label{xovo}}
\end{figure}

Now we want to return to the puzzling observation that  a charge transport occurs even without external bias. To understand the conditions under which such an effect occurs we have to clarify the dependencies on the initial kinetic and potential energy and the geometric configurations. The length $L$ of the nanoshafts determines only the time scale $\omega$ and scales out here. Therefore the distance between the nanoshafts $\Delta x$ in the row is the only geometric parameter left. Further, the initial kinetic and potential energy we determine by the elongation $x_0$ and speed $v_0$ of the leftmost nanoshaft while all other nanoshafts are initially not displaced and at rest. In figure \ref{xovo} we plot the dependence of the reached current on the initial displacement and the initial velocity. Using dimensionless units for the velocity $v*=v/\omega l_0$  and the displacement $x_0*=x_0/l_0$ the initial energy is $E_{\rm tot}=m\omega^2 l_0^2 (v_0*^2+x_0*^2)/2$ where the length unit is $l_0$. We see in the right plot of figure \ref{xovo} that the current is increasing quadratically with the initial velocity, i.e. the current is proportional to the initially deposited kinetic energy. This initial kinetic energy can be equivalently realized by thermal motion and characterized by a temperature. In the simulation described so far we bend the leftmost nanoshaft sufficiently close to the electrode such that the shuttling can happen even without initial kinetic energy. If we bend it less we obtain a threshold in kinetic energy analogously to the thermal threshold in Brownian motors \cite{HTB90}. 
The dependence of the current on the initial position  is seen in the left plot of figure \ref{xovo}. We have chosen an initial velocity below the threshold which shows that a threshold in the position or potential energy has to be overcome in order to create the shuttling current. We note that the shuttling transport develops only if we start with asymmetric initial conditions. Otherwise counter-oscillations of the left and right half of shafts block any transport. Therefore the symmetry breaking is due to initial conditions and this model  illustrates an electronic ratchet effect with thresholds in kinetic or potential energy corresponding to a high system temperature or initially elongation being $m \omega^2 \Delta x^2/2=281$meV for the here used example of quarterthiophene.

In summary we have found that a chain of perpendicularly arranged coupled and chargeable nanoshafts show a shuttling transport of charges. As a surprising effect it turns out that a finite current is established already without external bias only due to the initial asymmetric deformation of the nanoshafts.
The effect observed here is reminiscent of the ratchet effect. Even in the absence of a net macroscopic force or noise a current can be generated. This resembles a lot the prominent motor proteins found inside cells. There, proteins such as Mysosin and Kynesin use chemical energy which is gained from hydrolysis of ATP into ADP to move along asymmetric pathways transporting vesicles inside cells, contracting muscles and being important in the process of cell division \cite{JAP97,AH02}.

%\medskip

This work was supported by 
%research plans MSM 0021620834 and 
%No. AVOZ10100521, by grants GA\v{C}R 202/07/0597 and 202/06/0040 
%and GAAV 100100712 and IAA1010404, by PPP project of DAAD, by 
DFG 
Priority Program 1157 via GE1202/06 and the BMBF and by European 
ESF program NES.

\vspace{-3ex}

%\bibliography{kmsr,kmsr1,kmsr2,kmsr3,kmsr4,kmsr5,kmsr6,kmsr7,delay2,spin,gdr,refer,sem1,sem2,sem3,micha,genn,solid,shuttling}
%\bibliographystyle{prsty}

\end{document}